\pdfoutput=1
\documentclass[preprint,
		preprintnumbers,
		amsmath,
		amssymb,
		floatfix,
		nofootinbib]{revtex4}
\usepackage{epsf}
\usepackage{graphicx}
\usepackage{slashed}
\usepackage[textsize=tiny]{todonotes}
\usepackage{subfigure}
\usepackage{float}
\usepackage{color}
\usepackage[italicdiff]{physics}
\usepackage{textcomp}

\graphicspath{ {./figures/graphs/}{./figures/} }
\usepackage{import}
\usepackage{xifthen}
\usepackage{pdfpages}
\usepackage{transparent}
\usepackage{float}
\newcommand{\incfigwid}[2]{%
	\includegraphics[width=#2\columnwidth]{#1}
}
\newcommand{\incsubgraph}[1]{%
	\includegraphics[width=.63\columnwidth]{#1}
}

\newcommand{\A}{\mathcal{A}}
\newcommand{\M}{\mathcal{M}}
\renewcommand{\vec}[1]{\vb*{\vb{#1}}}
\newcommand{\xhat}{\vb{\hat{x}}}
\newcommand{\yhat}{\vb{\hat{y}}}
\newcommand{\zhat}{\vb{\hat{z}}}
\newcommand{\Day}{\Omega_{\oplus}}
\newcommand{\xixy}{\sqrt{\xi_x^2 + \xi_y^2}}

\begin{document}
\title{Breaking of Lorentz Invariance in Electron-Proton Parity Violation}
\author{Alexander Michel and Marc Sher}\email[] {asmichel@email.wm.edu, mtsher@wm.edu}

\affiliation{
Physics Department, William \& Mary, Williamsburg VA 23187
\vspace*{.5in}}

\date{\today}

\begin{abstract}
A popular framework for exploring Lorentz violation is the Standard Model
Extension. This extension contains a large number of parameters that can
be bounded in various experiments. Most studies, however, focus on the
fermion or photon sector. Here, we consider Lorentz violation in the weak
vector boson sector. The strongest bounds come from measurements of the
asymmetry in M\o ller scattering. We study the bounds that can be obtained from
future measurements of the parity violating asymmetry in deep inelastic
electron-proton scattering at the EIC, the LHeC and the FCC-eh.  For the
FCC-eh, the bounds could be substantially improved over current bounds by including timing information.
\end{abstract}
\maketitle

\section{Introduction}
While the Standard Model is successful at describing physical phenomena at the
highest energies measured, it is expected to break down when gravitational
effects are no longer negligible. Since many models of quantum gravity are
nonlocal, one might expect Lorentz invariance to be violated at a high energy
scale. While it is expected that this high energy scale would be the Planck
scale, requiring very high precision studies, larger effects are possible, and
the question of Lorentz violation is one that should be probed experimentally.

In general, it is difficult to write the most general Lorentz violating theory.
Even the meaning of a Lagrangian becomes uncertain in such a theory. An
extremely useful approach was developed years ago by Colladay and Kosteleck\'y
\cite{Colladay:1996iz,Colladay:1998fq,Kostelecky:2002hh,Kostelecky:2003fs}, who constructed the
Standard Model Extension (SME). This model is based on the Standard Model, but
adds Lorentz violating terms which satisfy the Standard Model gauge symmetry and
have dimension less than or equal to four. The extra terms also are invariant
under observer Lorentz transformations, i.e.\ all Lorentz indices must be
contracted and the physics does not depend on the choice of coordinates. The
Lorentz violation is also independent of position and time, so that energy and
momentum are conserved. The model contains a large number of parameters that can
be experimentally constrained, and there are many hundreds of papers studying
constraints on these parameters (an extensive list, updated to 2016, can be
found in Ref.\ \cite{Kostelecky:2008ts}).

\begin{figure}[h]
\centering
	\subfigure[{\label{pzg}}]
	{ \(\vcenter{\hbox{\incfigwid{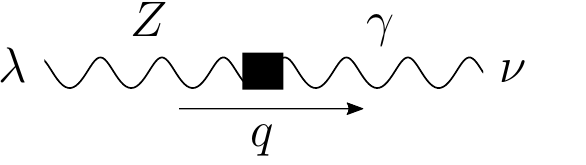}{.25}}} = -2i\tan\theta_W
	(k_W)_{\kappa\lambda\mu\nu} q^\kappa q^\mu\) } 
	
	\subfigure[{\label{pzz}}]{\(\vcenter{\hbox{\incfigwid{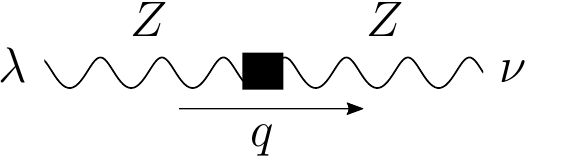}{.25}}} =
	 -2i(1-\tan^2\theta_W) (k_W)_{\kappa\lambda\mu\nu} q^\kappa
	q^\mu\)}
	\caption{\label{ins} Propagator insertions afforded by Eq.~\ref{lagr}.}
\end{figure}

The vast majority of these studies have involved long-lived particles, and only
a few have involved, for example, heavy scalar and vector bosons
\cite{Altschul:2006uw,Anderson:2004qi,Iltan:2003we,Iltan:2003nq,Iltan:2004qp,Aranda:2013cva,Aranda:2013vda,Noordmans:2013xga,Noordmans:2014hxa,Dijck:2013dza,Vos:2015fqi}.
Most of these involve fermions, studying, for example, beta-decays. Recently,
Fu and Lehnert \cite{Fu:2016fmf} considered bounds on parameters that only
involve gauge bosons. They studied bounds involving internal $Z$ boson lines in
electron-electron scattering experiments searching for parity violation, and
found that bounds from the E158 experiment at SLAC improved previous bounds
by two orders of magnitude. The parameters they considered arise from the
Lorentz violating terms in the SME
\begin{equation}
	\mathcal{L} = -\frac{1}{4}(k_B)_{\kappa\lambda\mu\nu}
	B^{\kappa\lambda}B^{\mu\nu} -\frac{1}{2}(k_W)_{\kappa\lambda\mu\nu}\tr
	(W^{\kappa\lambda}W^{\mu\nu})\,.
\end{equation}
The coefficients $k_B$ and $k_W$ are real and dimensionless. They have the
symmetries of the Riemann tensor and a vanishing double trace, so there are 19
coefficients each. These are CPT even; we will not discuss CPT odd terms since
they are associated with negative energy contributions. Fu and Lehnert also
discuss the fact that a relevant term in the Higgs kinetic energy SME
Lagrangian will not have an effect on the results, and we will ignore that
here. Writing the above Lagrangian term in terms of the photon and $Z$ gives
\begin{equation}
\begin{aligned}
	\mathcal{L} = &-\frac{1}{4}(k_B\cos^2\theta_W +
	k_W\sin^2\theta_W)_{\kappa\lambda\mu\nu}F^{\kappa\lambda}F^{\mu\nu} \\
	& - \frac{1}{4}(k_W\cos^2\theta_W +
	k_B\sin^2\theta_W)_{\kappa\lambda\mu\nu}Z^{\kappa\lambda}Z^{\mu\nu} \\
	& - \frac{1}{4}\sin 2\theta_W (k_W -
	k_B)_{\kappa\lambda\mu\nu}F^{\kappa\lambda}Z^{\mu\nu}
\end{aligned}
	\label{lagr}
\end{equation}
Here, $F^{\mu\nu} = \partial^\mu A^\nu - \partial^\nu A^\mu, Z^{\mu\nu} =
\partial^\mu Z^\nu - \partial^\nu Z^\mu$. The first term in this expression
deals with Lorentz violation in the QED sector. This is very strongly
constrained by many experiments and is completely negligible, so one can set
$k_B = -\tan^2\theta_W k_W$. The resulting Feynman rules are given in Figure
\ref{ins}. It is the $k_W$ coefficients that Fu and Lehnert bound from
considerations of parity violation in electron-electron scattering.
In this paper, we extend the work of Fu and Lehnert to include parity violation
in electron-proton scattering. This has been measured precisely at Qweak, and
will also be measured at the Electron-Ion Collider (EIC)\cite{Accardi:2012qut,Aschenauer:2014cki,Abeyratne:2015pma}, Large Hadron-electron
Collider (LHeC)\cite{AbelleiraFernandez:2012cc} and eventually at the Future Circular Collider (FCC-eh)\cite{Abada:2019lih}.  There have been studies
of effects of Lorentz violation in deep inelastic
scattering\cite{Kostelecky:2016pyx, Lunghi:2018uwj,Kostelecky:2018yfa}, but
these all consider the quark sector coefficients on the scattering. We
considered the pure gauge sector coefficients. The model and calculations are
presented in Section 2 and the results are discussed (for each proposed future
experiment) in Section 3, as are our conclusions.
\section{Electron-proton Scattering}
In parity violation in elastic $e p$ scattering, such as at Qweak, the
Lorentz violating effects are proportional to $Q^2/M_Z^2$, which is very small.
This suppression makes it impossible to strengthen existing bounds on $k_W$.

\begin{figure}[H]
\centering
	\subfigure[{\label{qmg} $i\M_\gamma$}]{\incfigwid{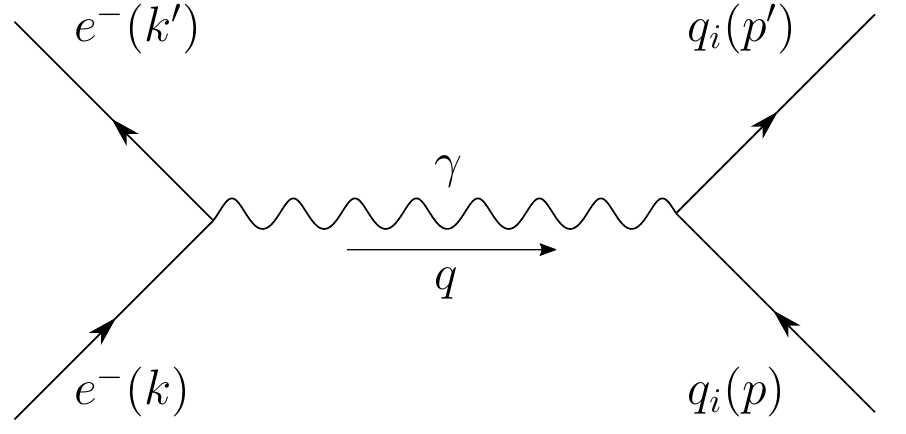}{.43}}
	\hfill
	\subfigure[{\label{qmz} $i\M_Z$}]{\incfigwid{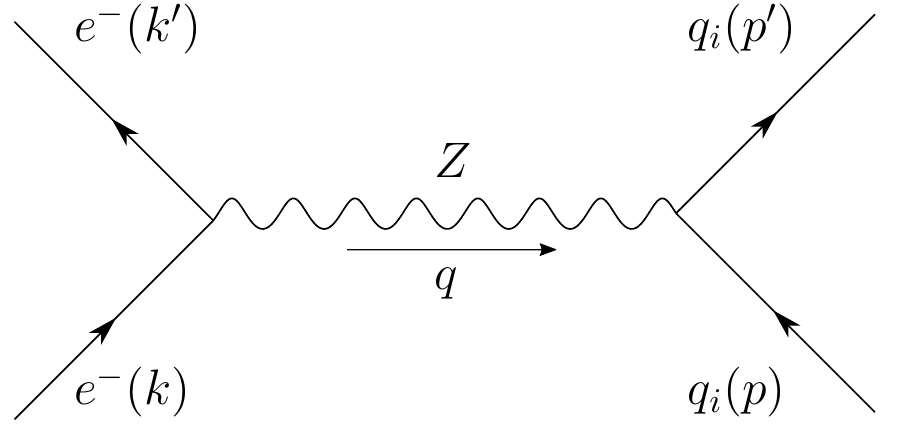}{.43}}
	\hfill
	\subfigure[{\label{qmgz} $i\M_{\gamma Z}$}]{\incfigwid{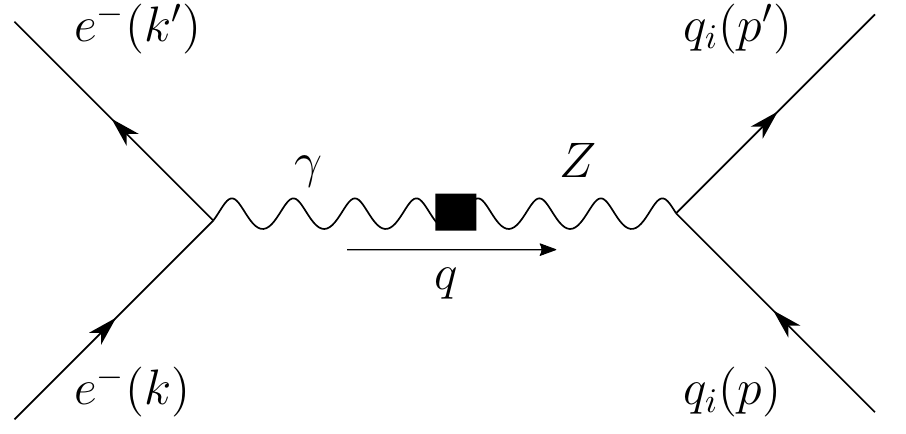}{.43}}
	\hfill
	\subfigure[{\label{qmzg} $i\M_{Z \gamma}$}]{\incfigwid{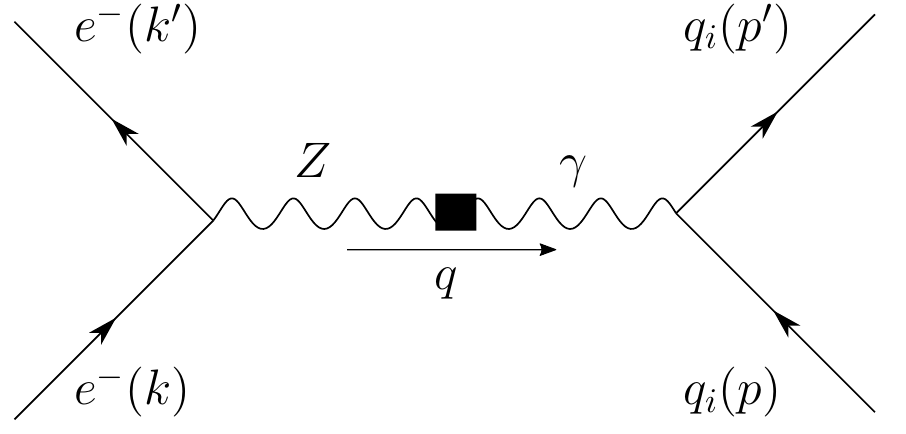}{.43}}
	\hfill
	\subfigure[{\label{qmzz} $i\M_{ZZ}$}]{\incfigwid{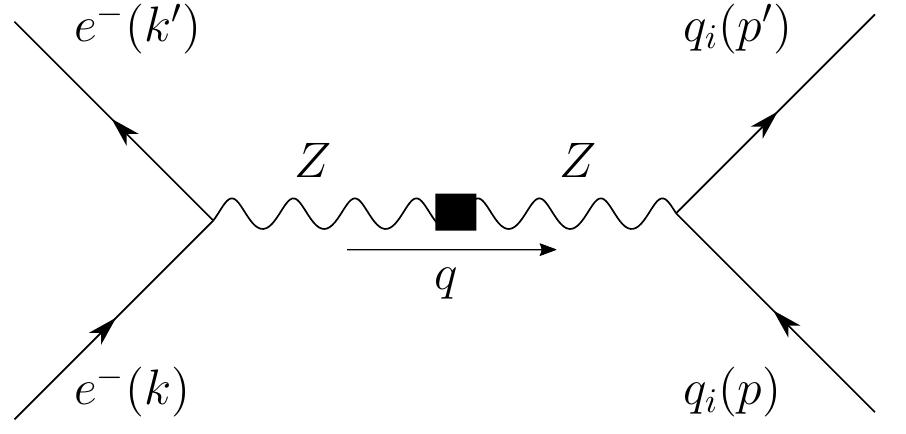}{.43}}
	\hfill
	\caption{\label{eqeq} Diagrams contributing to the process $e^-q_i\to
	e^-q_i$.}
\end{figure}

We instead consider inelastic $e^-p\to e^-X$ scattering. Only energies high
enough for the parton model to hold are considered. The $e^-p\to e^-X$ cross
section in terms of the $e^-q_i\to e^-q_i$ quark subprocess cross section is
\begin{equation}
	\qty(\frac{\dd[2]{\sigma}}{\dd{E'}\dd{\Omega}})_{ep\to eX} =
	\sum_{i}\int_0^1\,\dd{x}f_i(x)\qty(\frac{\dd[2]{\sigma}}{\dd{E'}
	\dd{\Omega}})_{eq_i\to eq_i}\,.
\label{parton}
\end{equation}
The functions $f_i(x)$ are the usual parton distribution functions
(PDFs) and the sum over $i$ ranges over all quark flavors. The diagrams
contributing to the quark subprocess are in Figure \ref{eqeq}.
We denote by $\abs{\M_{i,h}}^2$ the squared sum of the diagrams in Figure
\ref{eqeq}, where $h=L,R$ is the helicity of the incoming electron and $i$ is
the quark flavor. This matrix element can be written \[\abs{\M_{i,h}}^2 =
\abs{\M_{i,h}}_0^2 + \delta\abs{\M_{i,h}}^2\,,\] where $\abs{\M_{i,h}}_0^2$
contains the usual standard model contribution and $\delta\abs{\M_{i,h}}^2$ is
the Lorentz violating correction. The explicit form of $\delta\abs{\M_{i,h}}^2$
is rather long, so we have placed it in Appendix A.

The primary observable is the asymmetry
\begin{equation}
\A \equiv \frac{d\sigma_L(ep\to eX)-d\sigma_R(ep\to eX)}{d\sigma_L(ep\to
eX)+d\sigma_R(ep\to eX)}\,. 
\end{equation}
Expanded to leading order in $k_W$ in the lab frame, the asymmetry is of the
form $\A = \A_0 + \delta\A$ where $\A_0$ is the Standard Model result and
$\delta\A$ is the Lorentz violating correction proportional to $k_W$.

In order to cast $\delta\A$ into a form suitable for numerical analysis, we
employ a parameterization of $k_W$ due to Fu and Lehnert \cite{Fu:2016fmf} in
terms of a dimensionless vector $\vec{\xi}=(\xi_x,\xi_y,\xi_z)$.  This is a substantial simplification of the analytical calculation, which 
reduces the 19-dimensional parameter space to a 3-dimensional parameter space.   It seems reasonable, since time and space coordinates do not mix, 
but there is no general argument as to why these three parameters should dominate the effects.    Fu and Lehnert then define $\xi^\mu \equiv (0,\vec{\xi})^\mu$ and $\zeta^\mu \equiv
(1,\vec{0})^\mu$ we can write
{
\newcommand{\comu}[2]{\xi^{\{#1}\zeta^{#2\}}}
\begin{equation}
\begin{gathered}
(k_W)^{\kappa\lambda\mu\nu} =
\frac{1}{2}\left[g^{\kappa\mu}\comu{\lambda}{\nu} -
g^{\kappa\nu}\comu{\lambda}{\mu} + g^{\lambda\nu}\comu{\kappa}{\mu} -
g^{\lambda\mu}\comu{\kappa}{\nu}\right]\label{comu}\,,\\
\qq{where}\xi^{\{\mu}\zeta^{\nu\}} \equiv \frac{1}{2} \left(\xi^\mu\zeta^\nu +
\xi^\nu\zeta^\mu\right)
\end{gathered}
\end{equation}
}%
This parameterization reduces our problem to the bounding of $\vec{\xi}$, which
now expresses the entirety of the Lorentz violation effect.

One needs to consider the coordinate system in order to define $k_W$ (and
hence $\vec{\xi}$). The canonical choice is sun-centered celestial equatorial
coordinates \cite{Kostelecky:2002hh}. Consequently, the calculation of
$\delta\A$ will require all momenta to be expressed in this frame. To change to
these coordinates from those usually employed in scattering
calculations\footnote{E.g.\ coordinates chosen so that the plane of interaction
is spanned by $\zhat$ and either of $\xhat,\yhat$.} one must first transform to
the coordinate system with $\xhat$ pointing south, $\yhat$ pointing east, and
$\zhat$ normal to the Earth's surface. Changing to this reference frame will
involve rotations that depend on the azimuthal scattering angle $\phi$ and the
angle $\beta$ of the electron beam east of south. The rotation
\begin{equation}
\begin{pmatrix}
\cos\chi\cos\Day t & -\sin\Day t & \sin\chi\cos\Day t \\
\cos\chi\sin\Day t & \cos\Day t & \sin\chi\sin\Day t \\
-\sin\chi & 0 & \cos\chi
\end{pmatrix}
\label{dayrotation}
\end{equation}
then transforms to the sun-centered celestial equatorial coordinate system
\cite{Kostelecky:2002hh}. Here, $\Day=2\pi/(23\text{ h }56\text{ min})$ and
$\chi$ is the colatitude of the lab. This last rotation need not include a
boost, since the Earth's motion is completely nonrelativistic.

Notice that the only 3-vectors appearing in the problem are the incoming
(outgoing) electron momenta $\vec{k}$ ($\vec{k'}$) and the parton momentum
$\vec{p}$. Then since $\delta\A$ is a scalar quantity proportional to $k_W$, it
must be a linear combination of the dot products $\vec{\hat{k}}\cdot\vec{\xi}$,
$\vec{\hat{k'}}\cdot\vec{\xi}$, and $\vec{\hat{p}}\cdot\vec{\xi}$.
It is then straightforward to show that
\begin{equation}
\begin{aligned}
\vec{\hat{p}}\cdot\vec{\xi}&=-\vec{\hat{k}}\cdot\vec{\xi} \\
\vec{\hat{k}}\cdot\vec{\xi} &= f_1\cos(\Day t - \delta_1)\xixy\qq{+}
\text{constant} \\
\vec{\hat{k}'}\cdot\vec{\xi} &= \big(f_2\cos(\Day t - \delta_2) + f_3\cos\Day
t\big)\xixy\qq{+}\text{constant} \\
\end{aligned}
\label{dots}
\end{equation}
The parameters used in the above expressions are given in Table 1. The overall
additive constants we have chosen to omit above are time independent and
therefore not easily disentangled from Standard Model effects. Fu and Lehnert
also find a dependence on $\sqrt{\xi^2_x+\xi^2_y}$. The $\xi_z$ piece is
not sensitive to the time dependence since the $z$-direction is perpendicular
to the Earth's surface, and the azimuthal dependence vanishes, as they show, in
the ultrarelativistic limit.
\begin{table}
\centering
{
\renewcommand{\hline}{}
$\begin{array}{||c | c ||}
\gamma & \arctan(\tan\theta\sin\phi) \\
\delta_1 & -\arctan({\tan\beta}/{\cos\chi}) \\
\delta_2 & -\arctan({\tan(\beta-\gamma)}/{\cos\chi}) \\
f_1 & \sqrt{1-\sin^2\beta\cos^2\chi} \\
f_2 &
\sqrt{1-\sin^2\theta\cos^2\phi}\sqrt{1-\sin^2(\beta-\gamma)\cos^2\chi} \\
f_3 & \abs{\sin\theta\cos\phi\sin\chi} \\
\end{array}$
}
\caption{\label{params} Explicit forms of the parameters in Eq.~\ref{dots}.
Recall that $\beta$ is the beam angle east of south, $\chi$ is the colatitude
of the beam, and $\theta$ ($\phi$) is the polar (azimuthal) scattering angle of the scattered electron.}
\end{table}
Using Eq.~\ref{dots} we can cast the correction into its final form
\begin{equation}
\delta\A = A(\theta,\phi)\cos\big(\Day t - \delta(\theta,\phi)\big)\xixy
\qq{+}\text{constant}
\label{deltaA}
\end{equation}
The computation of $\delta$ and $A$ must be done numerically as a function of
$\theta$, all the parameters in Table 1, and the beam energies associated
with the given experiment. This is the subject of the next section.

\section{Results and Conclusions}

The calculation is now straightforward. One uses Eq.~A3, with $F$ (the proton
energy) replaced by the quark energy, $xF$, and then kinematically replaces
$E^\prime$ with \[\frac{2xEF}{E+xF-(E-xF)\cos\theta}\,.\] The $c_i^{V,A}$ are
then input into Eq.~A2 and then the matrix elements in Eq.~A4 are summed. Note
the factor of $h$ in the matrix elements---this is $-1$ for left-handed
electrons and $+1$ for right-handed electrons. The dot products involving
$\vec{\xi}$ are in Eq.~7. Then one integrates over the parton distributions and
the asymmetry is determined. The final form is in Eq.~8, and we just need to
determine $A(\theta,\phi)$.

The resulting asymmetry depends on the colatitude, $\chi$, which we will take
to be 45 degrees (this is fairly close to all current and future laboratories).
It depends on $\beta$, the angle of the beam east of south, which will, in due
course, be known for any collider. The only remaining variables are the polar
and azimuthal angle of the scattered electron. One could, of course, integrate
over these variables, but it is useful to see the dependence explicitly.  

\begin{figure}
\centering
\makebox[\textwidth][c]{%
\subfigure[\label{fccbeta} Fixed
$\beta=45$\textdegree]{\incsubgraph{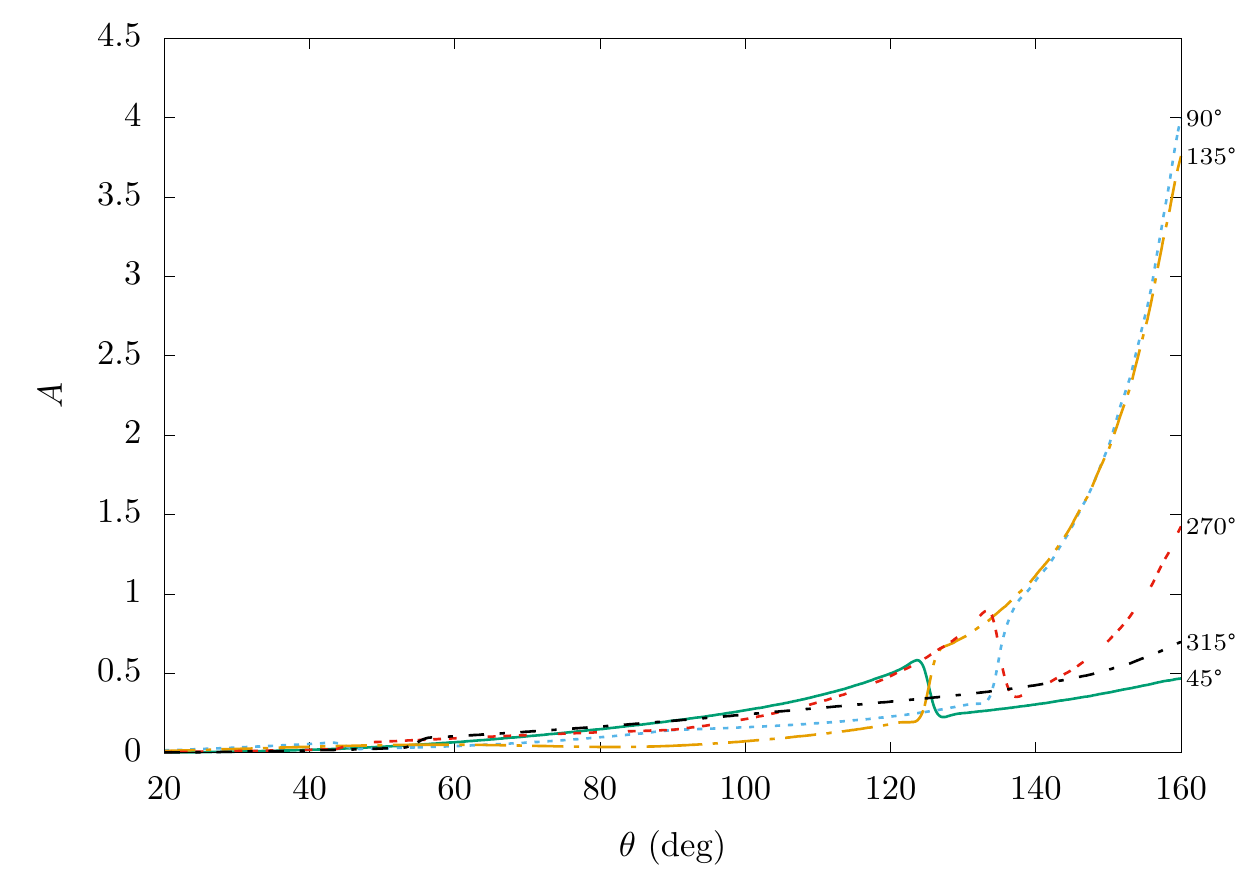}}
\hfill\hspace{-2em}
\subfigure[\label{fccphi} Fixed
$\phi=60$\textdegree]{\incsubgraph{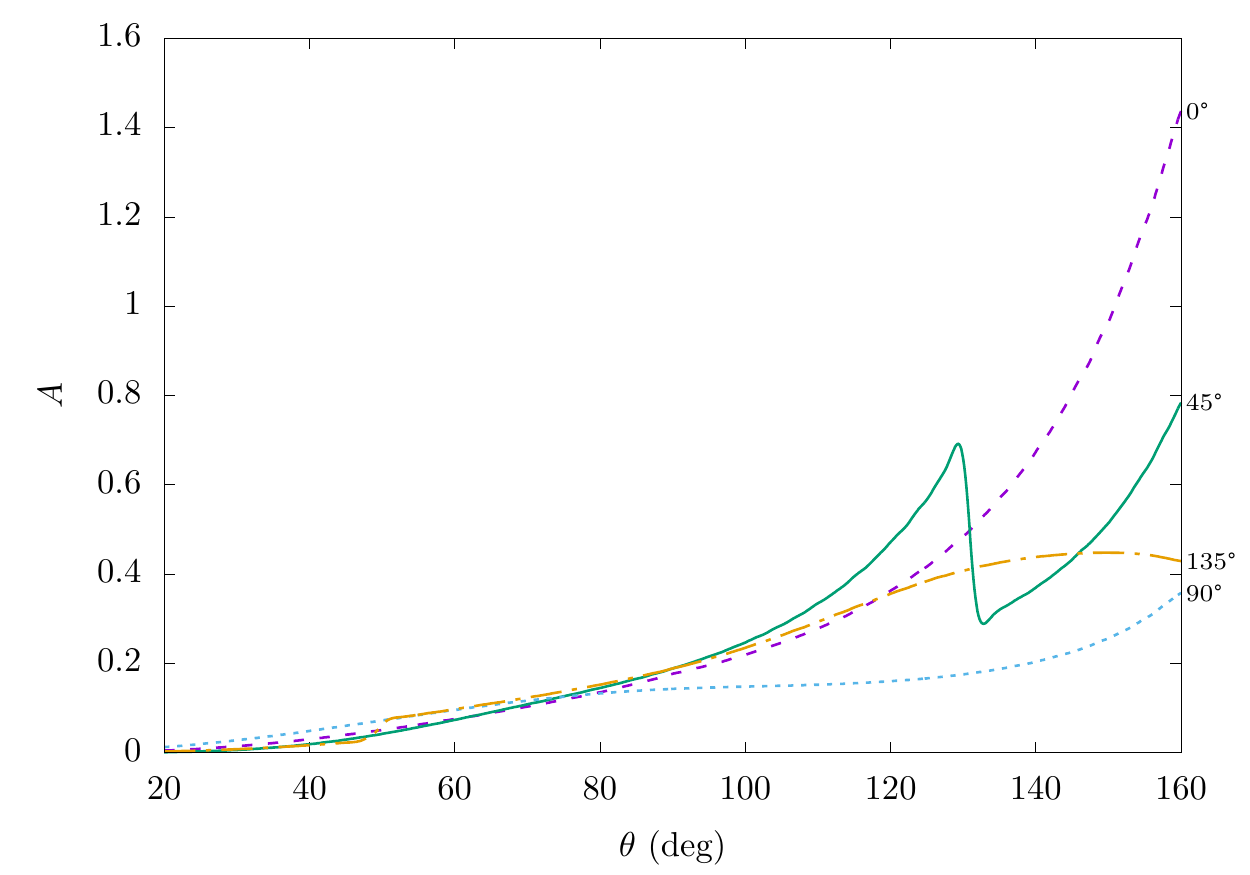}}}
\caption{\label{fcceh} Graphs of $A(\theta,\phi)$ for FCC-eh energies. In (a),
$\beta$ is fixed to $45$\textdegree\ while $\phi$ takes on the values on the
right of the graph. The reverse is true for (b).}
\end{figure}

Our results for the FCC-eh are in Figure \ref{fcceh}. In Figure \ref{fccbeta},
we show the value of A (in Equation \ref{deltaA}) as a function of $\theta$ for
various values of $\phi$. Here, we have set $\beta = 45$ degrees.    Note that
the Lorentz violating corrections are peaked in the backward direction. In
Figure \ref{fccphi}, we show the value of A as a function of $\theta$ for
various values of $\beta$, setting $\phi = 60$ degrees. If one integrates over
the solid angle in the backward hemisphere and divides by $2\pi$, one can get
an ``average'' value of $A$.  Typical values of $A$ are 0.3-0.7, depending on
$\beta$, and thus the result in Equation 8 is this value times $\sqrt{\xi^2_x +
\xi^2_y}$ times an oscillation with a period of the Earth's rotational period.

\begin{figure}
\centering
\makebox[\textwidth][c]{%
\subfigure[\label{lhecbeta} Fixed
$\beta=45$\textdegree]{\incsubgraph{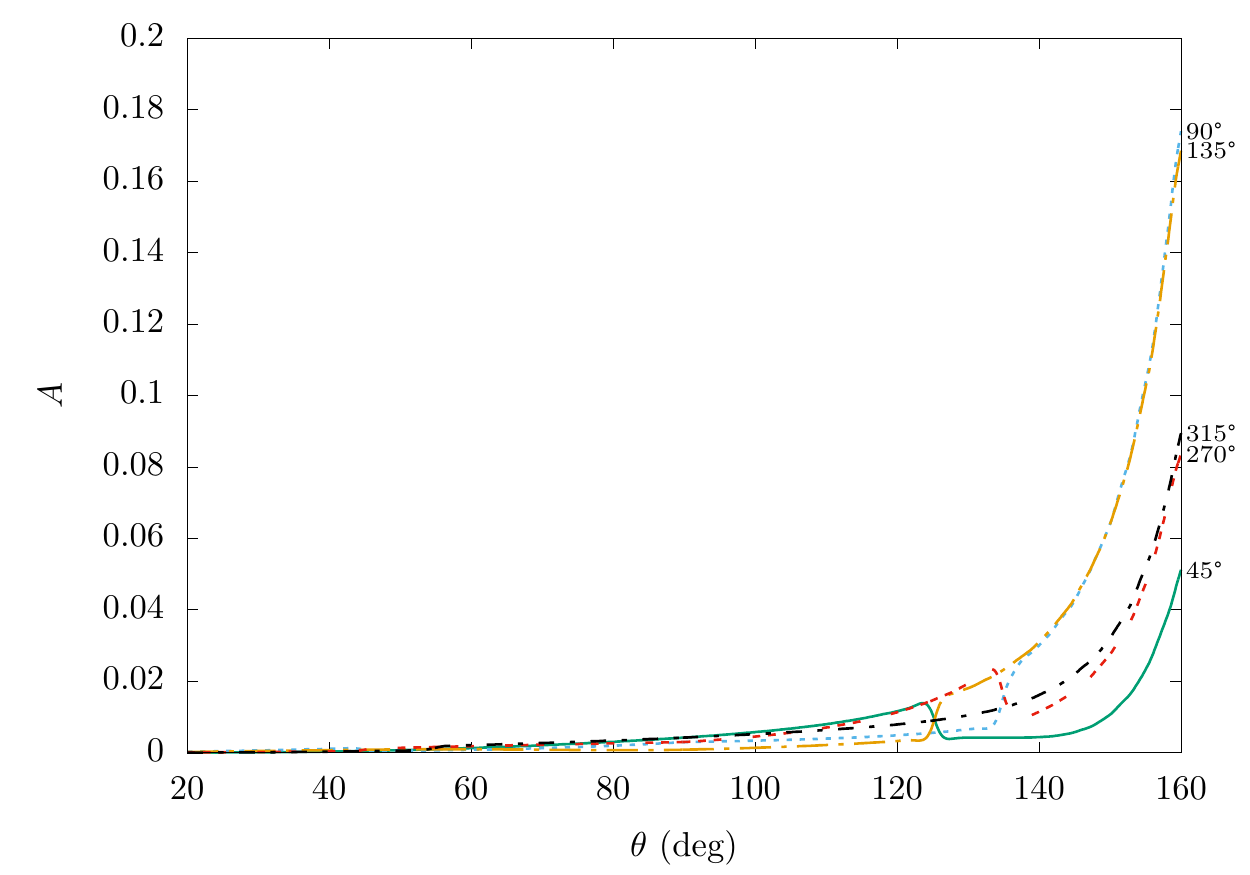}}
\hfill\hspace{-2em}
\subfigure[\label{lhecphi} Fixed
$\phi=60$\textdegree]{\incsubgraph{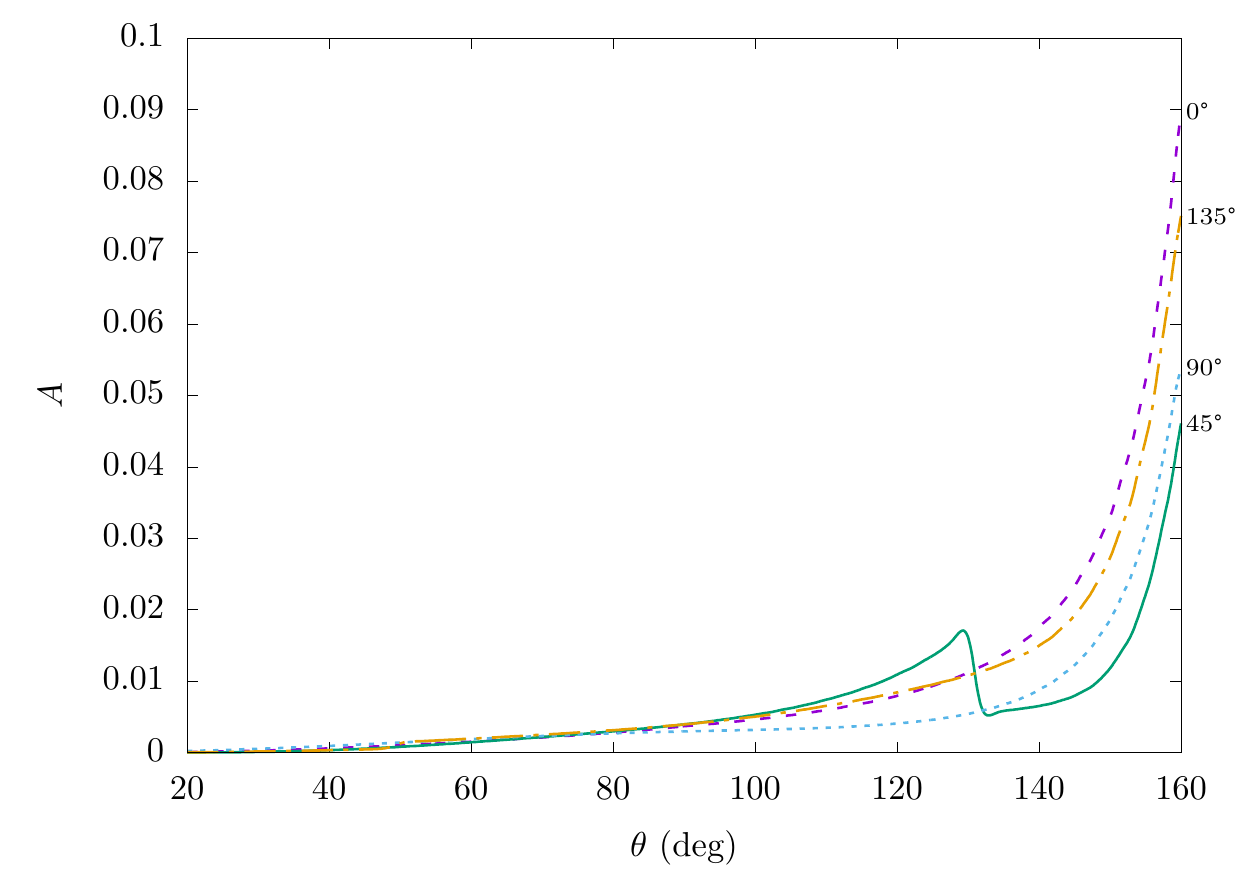}}}
\caption{\label{lhec} Graphs of $A(\theta,\phi)$ for LHeC energies.}
\end{figure}

% \begin{figure}
% \centering
% \makebox[\textwidth][c]{%
% \subfigure[\label{eicbeta} Fixed
% $\beta=45$\textdegree]{\incsubgraph{EIC-beta45}}
% \hfill\hspace{-2em}
% \subfigure[\label{eicphi} Fixed $\phi=60$\textdegree]{\incsubgraph{EIC-phi60}}}
% \caption{\label{eic} Graphs of $A(\theta,\phi)$ for EIC energies.}
% \end{figure}

In Figure 4, we have given the results for the LHeC. The general structure
is similar to the FCC-eh, but the amplitude is roughly an order of magnitude
smaller. We will not show results for the EIC---the structure is similar but
the amplitude is 2-3 orders of magnitude smaller.   The reason that the LHeC and EIC have much smaller effects is entirely due to the
center of mass energy of the colliders.   The LHeC has a proton beam that has an order of magnitude less energy, and the EIC has an even smaller energy.

In the work of Fu and Lehnert \cite{Fu:2016fmf}, the value of $\sqrt{\xi^2_x +
\xi^2_y}$ was bounded by studying the E158 experiment at SLAC
\cite{Anthony:2003ub, Anthony:2005pm}, which studied M\o ller scattering. The
bound arose by requiring that the amplitude of the fluctuation be smaller than
the statistical uncertainty in the E158 experiment, and gave an upper bound of
$3.4\times 10^{-7}$. They noted that the upcoming MOLLER experiment at
Jefferson Lab would, using the same argument, be able to improve this by a
factor of $1.4$. They also pointed out that the different beam directions
might provide access to different components of $(k_W)_{\kappa\lambda\mu\nu}$,
thus even if the bound could not be improved, it would be valuable to search
for Lorentz violation.   This is already a low bound and it might seem difficult for the FCC-eh 
to do much better.

The Fu-Lehnert analysis only considered the amplitude of the oscillation---the
actual rotation period of the Earth did not enter into the analysis. This is
not surprising, since the timing of events for E158 are not available. If
such timing is included for the FCC-eh, then it might be possible to perform a
fit including the explicit time-dependence and thus improving the bound.
Even if one does not do this, it is likely that a good bound could be obtained.
While we do not know the estimated uncertainties in DIS parity violation at the
FCC-eh, it has been reported\cite{Abada:2019lih} that DIS at the FCC-eh will be
sensitive to $\sin^2\theta_W$ to an accuracy of $0.001$. This is a factor of
four better than the E158 experimental uncertainty, and thus it is plausible
that the FCC-eh will improve the bound substantially.  At the LHeC, the size of the effect is an order of magnitude smaller, but a comparable bound might still be reached.

Even if the bound obtained by E158 cannot be reached, one must remember that
the Fu-Lehnert parameterization of Eq.~5, while quite reasonable, essentially
considers a three dimensional slice of the 19-dimensional space of
$(k_W)_{\kappa\lambda\mu\nu}$. Thus, a fit including to a potential time-dependence in future
deep inelastic experiments could prove worthwhile.

In future deep inelastic parity violation experiments, the values of the
latitude and electron beam direction will be fixed, and it will be
straightforward to use our results to find the polar and azimuthal angular
dependence of the Lorentz violating effect.    By fitting to a sinusoidal dependence with a period of the Earth's rotation, one will be able
to bound $\sqrt{\xi^2_x + \xi^2_y}$, if the Fu-Lehnert parametrization is
adopted.    But even without that parametrization, a search would be
worthwhile, since it is sensitive to the entire 19-dimensional space of
$(k_W)_{\kappa\lambda\mu\nu}$.

\section*{Acknowledgements}
We are grateful to Wouter Deconinck, David Armstrong, Jianwei Qiu, Michael
Kordosky and Chris Carone for useful discussions. We thank the National
Science Foundation for support under Grant PHY-1819575.

This work was performed in part using computing facilities at the College of
William and Mary which were provided by contributions from the National Science
Foundation, the Commonwealth of Virginia Equipment Trust Fund and the Office of
Naval Research.

\appendix
\section{Computation of \boldmath{$\delta\abs{\M_{i,h}}^2$}}
The terms in
{ 
\newcommand{\temm}[2]{\M_{#1}^*\M_{#2}}
\[\delta\abs{\M_{i,h}}^2 = 2\Re\big\{\temm{\gamma}{\gamma Z} +
\temm{\gamma}{Z\gamma} + \temm{\gamma}{ZZ} + \temm{Z}{\gamma Z} +
\temm{Z}{Z\gamma} + \temm{Z}{ZZ}\,\big\}\]
}%
all have the same trace structure. Each has a factor of
$(k_W)^{\kappa\lambda\mu\nu}(k^\prime_\kappa-k_\kappa)(k^\prime_\mu-k_\mu)$
times either of 
{
\renewcommand{\tr}{\trace}
\begin{equation}
\tr[\slashed{k'}\gamma^\lambda\slashed{k} \gamma_\rho]
\tr[\slashed{p'}\gamma^\nu\slashed{p}\gamma^\rho]\qq{or}
\tr[\gamma^5\slashed{k'}\gamma^\lambda\slashed{k}\gamma_\rho]
\tr[\gamma^5\slashed{p'}\gamma^\nu\slashed{p}\gamma^\rho]\,.
\label{traces}
\end{equation}
}%
Let $E$ be the incoming electron energy, $E'$ the scattered electron
energy, $\theta$ the scattering angle, and $F$ the incoming proton energy.
Then using the Fu-Lehnert parameterization one can write
$(k_W)^{\kappa\lambda\mu\nu}(k^\prime_\kappa-k_\kappa)(k^\prime_\mu-k_\mu)$
times either of the traces in Eq.~\ref{traces} as 
\begin{equation}Y^{V,A} = c_1^{V,A} \vec{k}\cdot\vec{\xi} + c_2^{V,A}
\vec{k}^\prime\cdot\vec{\xi} + c_3^{V,A} \vec{p}\cdot\vec{\xi}\,.\end{equation}
Here, $c_j^V$ and $c_j^A$ refer to the first and second traces in
Eq.~\ref{traces}, respectively. Abbreviating $c\equiv\cos\theta$ for notational
clarity, one finds
\begin{align}
c_1^V &= 8 F E^\prime[-5(1-c^2)EE^\prime F +
4(c-1)E^3-2(1+c)E^2((c-1)E^\prime-F)+(c+1)^2E^{\prime 2}F]\nonumber \\
c_2^V &= 16 E F[(c-1)EE^\prime(2E^\prime+F)-(c+1)E^{\prime
2}((c-1)E^\prime+(c-2)F)-2E^2F]\nonumber \\
c_3^V &= 8(c-1)EE^\prime F[(7c+5)EE^\prime - (c+1)E^{\prime 2}-2E^2\nonumber \\
c_1^A&= 8(1-c)EE^\prime F[(1+c)EE^\prime + (1+c)E^{\prime 2} -4E^2]\nonumber \\
c_2^A&= 16(1-c)EE^\prime F(E-E^\prime)((1+c)E^\prime + E)\nonumber \\
c_3^A&= -8(1-c)^2E^2E^{\prime 2}(E + E^\prime)
\end{align}
This gives
\begin{align}
2\Re(M_\gamma^*M_{\gamma Z}) &= \frac{-Qe^4}{2\cos^2\theta_w t^2 (t-M_Z^2)}
(G_V Y^V - h G_A Y^A)\nonumber \\
2\Re(M_\gamma^* M_{Z\gamma} )&= \frac{Q^2e^4(g_V-hg_A)}{2\cos^2\theta_w t^2
(t-M_Z^2)} Y^V\nonumber \\
2\Re(M_\gamma^* M_{ZZ} )&=
\frac{Qe^4(1-\tan\theta_W)(g_V-hg_A)}{4\cos^2\theta_w\sin^2\theta_w t
(t-M^2_Z)^2} (G_V Y^V - h G_A Y^A)\nonumber \\
2\Re(M_Z^*M_{\gamma Z})&= \frac{(g_V-hG_A)e^4}{8\cos^4\theta_w\sin^2\theta_w t
(t-M^2_Z)} ((G_V^2+G_A^2)Y^V -2G_VG_Ah Y^A)\nonumber \\
2\Re(M_Z^*M_{Z\gamma})&=
\frac{-(g_V-hg_A)^2Qe^4}{8\sin^2\theta_w\cos^4\theta_w t (t-M^2_Z)^2} (G_VY^V -
hG_AY^A)\nonumber \\
2\Re(M_Z^*M_{ZZ})&=
\frac{-(g_V-hg_A)^2e^4(1-\tan\theta_w)}{16\sin^4\theta_w\cos^4\theta_w
(t-M^2_Z)^3}((G_V^2+G_A^2)Y^V -2G_VG_Ah Y^A)\nonumber \\
\end{align}
Here, $g_V$ and $g_A$  ($G_V$ and $G_A$) are the vector and axial vector couplings of the electron and quarks, respectively, and $Q$ is the quark charge.

\end{document}